\documentclass[doublecol]{epl2}
\usepackage{amsmath,amsfonts,amssymb}
\usepackage{graphicx}
\usepackage{algorithm}
\usepackage{algorithmic}
\newcommand{\arctanh}[1]{\mathrm{arctanh}#1}

\def\<{\langle}
\def\>{\rangle}
\def\Ham{\mathcal{H}}

\newcommand{\Jij}{J_{i,j}}

\def\qab{q_{\alpha \beta}}

\def\sia{s_i^\alpha}
\def\sib{s_i^\beta}
\def\sja{s_j^\alpha}

\def\mia{m_i^\alpha}
\def\mib{m_i^\beta}
\def\mja{m_j^\alpha}
\def\mjb{m_j^\beta}

\def\Cija{C_{ij}^\alpha}
\def\Cijb{C_{ij}^\beta}

\begin{document}
\title{Message passing and Monte Carlo algorithms: connecting fixed points with metastable states}
\author{A. Lage-Castellanos\inst{1} \and R. Mulet\inst{1} \and F. Ricci-Tersenghi\inst{2}}
\shortauthor{A. Lage-Castellanos \etal}
\institute{                    
  \inst{1} Department of Theoretical Physics, Physics Faculty,
  University of Havana, La Habana, CP 10400, Cuba. \\
  \inst{2} Dipartimento di Fisica, INFN -- Sezione di Roma 1 and CNR -- IPCF, UOS di Roma,
  Universit\`{a} La Sapienza, P.le A. Moro 5, 00185 Roma, Italy\\
}

\date{\today}

\abstract{Mean field-like approximations (including naive mean field,
  Bethe and Kikuchi and more general Cluster Variational Methods) are
  known to stabilize ordered phases at temperatures higher than the
  thermodynamical transition. For example, in the Edwards-Anderson
  model in 2-dimensions these approximations predict a spin glass
  transition at finite $T$. Here we show that the spin glass solutions
  of the Cluster Variational Method (CVM) at plaquette level do
  describe well actual metastable states of the system.  Moreover, we
  prove that these states can be used to predict non trivial
  statistical quantities, like the distribution of the overlap between
  two replicas. Our results support the idea that message passing
  algorithms can be helpful to accelerate Monte Carlo simulations in
  finite dimensional systems.
}

\maketitle

\section{Introduction}

Monte Carlo methods are the most celebrated and used  techniques to computationally explore the configuration space of Hamiltonian systems\cite{Newmanbook}. 
Unfortunately, in many practical cases, usually at very low temperatures or close to phase transitions,  the dynamics becomes very slow and the time needed to average the system diverges with the system size. 
The situation is specially frustrating when studying problems that are computationally demanding. In these cases it is natural to first try to understand the properties of the phase space that make the problems {\em hard} in the computational sense and then, with the help of this comprehension, to design efficient algorithms\cite{MezMon,MP2,KS2,Col}.

A very promising tool in this direction are message passing algorithms that are derived from an approximated free energy of the specific model of interest\cite{yedidia}. Up to now, most of the attention to this approach has been concentrated on Bethe-like approximations \cite{yedidia,KS2,KS,col2}. However, the applicability of this approximation is usually restricted to systems with very large loops, $\sim \log(N)$, where $N$ is the system size, but is of limited value to study finite dimensional systems.

A more sophisticated approach is the Cluster Variational Method (CVM)
\cite{tommaso,dual,GBPGF,zhou,pelizzola05} that in principle may consistently account for the presence of short loops in the model, providing also a more natural connection with MC methods in finite dimensional systems.

It is tempting to combine message passing and Monte Carlo techniques to exploit the potentialities of both approaches. In a recent paper, \cite{aurelien} this was done for the first time. In that contribution the standard Metropolis technique, was {\em guided} by the marginals estimated by a message passing algorithm defined on a proper tree-like structure. Yet, there is a lot of room for improvement. In particular, to use this, or similar techniques in finite dimensional systems. But to firmly progress in this direction it remains to understand what is the connection (if any) between the fixed points solutions of message passing techniques and Monte Carlo simulations in finite dimensional systems.

This is the main aim of this work. In what follows we present new data supporting that non-paramagnetic fixed points of plaquette-CVM are indeed connected with the configurational space explored by the Metropolis algorithm. These fixed points correspond to actual metastable states of the system and are a useful tool to extract non trivial information about the dynamics, for example the overlap between Monte Carlo replicas. We will use $\pm J$ Edwards-Anderson 2D model as proof of concept, but we expect that similar properties hold in other disordered models.

\section{Cluster Variational Method in Edwards-Anderson 2D}

The celebrated Edwards-Anderson model in statistical mechanics \cite{EA_original} is defined by a set $\sigma=\{s_1\ldots s_N\}$ of $N$ Ising spins $s_i=\pm 1$ placed at the nodes of a square lattice (in our case in two dimensions), and random interactions $\Jij = \pm 1$ at the edges, with a Hamiltonian
\begin{equation}\Ham (\sigma) = - \sum_{<i,j>} \Jij s_i s_j\end{equation}
where $<i,j>$ runs over all couples of neighboring spins. 

The direct computation of the partition function $Z$, or any marginal probability distribution like $p(s_i,s_j)=\sum_{\sigma \backslash s_i,s_j}P(\sigma)$, is  unattainable in general, and therefore approximations are required. Among all of them, we will explore the Cluster Variational Method (CVM), a technique that includes mean-field and Bethe approximations \cite{bethe} as particular cases, and was first derived by Kikuchi \cite{kikuchi} for the homogeneous system, and later extended to disordered models. In its modern presentation \cite{pelizzola05,yedidia}, it consists of replacing the exact (Boltzmann-Gibbs) distribution $P(\sigma)$, by a reduced set of its (approximated) marginals $\{b_R(\sigma_R)\}$ over certain degrees of freedom grouped in regions. With this reduction, the approximated free energy can be minimized in a numerically treatable manner. The consistency between the probability distributions of regions that share one or more degrees of freedom, is forced by Lagrange multipliers \cite{yedidia}. The latter are connected by self-consistent relations, that are solved by an iterative procedure, the so-called Generalized Belief Propagation (GBP). In what follows we use this approximation with the square plaquettes of the 2D lattice as the largest set of marginals considered. We skip the details and point the reader to \cite{GBPGF} where the precise form of these equations for the plaquette-CVM in EA 2D can be found. Other approaches, similar in spirit, have been followed in references \cite{chertkov,zhou,zhou2,zhou3}.

\section{Solutions of GBP}

When running GBP for the plaquette-CVM approximation in EA 2D we find a paramagnetic solution at high temperature, as expected. However, above $\beta_{\text{c}}\simeq 0.79$ ( below $T_{\text{c}}\simeq 1.26$ ) GBP finds, not one, but many fixed points with non-zero local magnetizations. Suggesting then, a transition from a paramagnetic to an spin glass phase \cite{GBPGF,average_mulet}. At still lower temperatures  (above $\beta_{\text{conv}}\simeq 1.2$) \cite{GBPGF}, GBP stops converging. The use of a provably-convergent method called Double Loop \cite{HAK03} showed that \cite{dual} in order to keep converging, the algorithm is set back to the paramagnetic solution. Is this region of  intermediate temperatures that will concentrate our attention here.

\begin{figure*}[!htb]
        \begin{center}
\includegraphics[width=0.25\textwidth, angle=270]{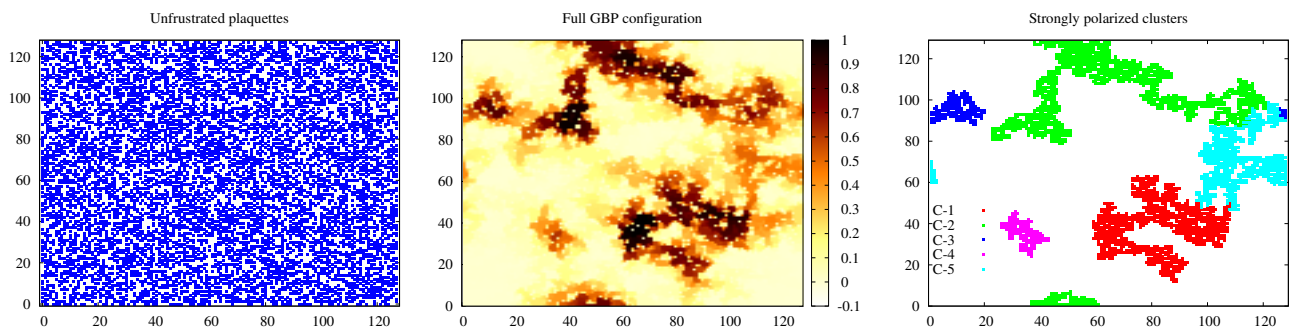} 
 \end{center}
\caption{For a $128\times 128$ EA2D system, the pattern of frustrated plaquettes (left panel) gives not obvious hint for the appearance of strongly magnetized regions (center panel) nor the distinction of clusters (right panel). In the clusters where the magnetization appears, there is a slightly higher concentration of non frustrated plaquettes ($53\%$ vs $49\%$ in the whole system).}
 \label{fig:clusters}
\end{figure*}

Already in references \cite{average_mulet} and \cite{zhou} it was noted that the non paramagnetic solutions have inhomogeneous magnetizations, not only in their sign as expected in a disordered system, but also in their spatial distribution: connected clusters of magnetized spins are surrounded by a sea of unmagnetized ones (see figure \ref{fig:clusters}).

\section{Relation to Monte Carlo}

\begin{figure}[!htb]
        \begin{center}
\includegraphics[scale=0.3, angle=0]{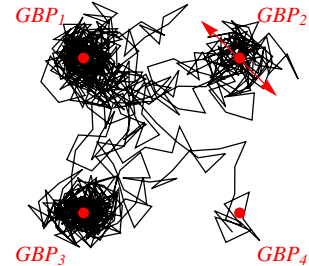} 
 \end{center}
\caption{Schematic representation of the Monte Carlo dynamics in the configurational space.  Most of the time MC is in the vicinity of one GBP solution.}
 \label{fig:confspace}
\end{figure}

Though GBP solutions are not thermodynamic states, we will show that Monte Carlo dynamics remain most of the time near the GBP solutions in the range of temperatures ($\beta_{\text{c}}-\beta_{\text{conv}}$), as schematized in figure \ref{fig:confspace}. 

A systematic approach (although heuristic and non exhaustive) to locate {\it all} GBP solutions is divided in two steps:  {\bf (i)} identify the clusters of connected and strongly magnetized spins, then {\bf (ii)} explore all possible combinations of orientations for those clusters. Locating the clusters starts from a given non paramagnetic solution of GBP equations (the reference GBP state) at the desired temperature. Then it iterates the following procedure, starting from $c=1$
\begin{enumerate}
 \item Take the most magnetized spin that do not belong to a cluster already defined, call its magnetization $m_c$
 \item Add to cluster $c$ all connected (nearby) spins that have a magnetization with modulus greater than $\theta |m_c|$, where $\theta$ is a not too small threshold parameter (we show results here with $\theta  = 0.8$, but other values produced equivalent results).
 \item If cluster $c$ is not in touch with any previously defined cluster, then go to 1 with $c=c+1$, else stop.
\end{enumerate}
The result of the procedure for a particular instance is depicted in the rightmost panel of figure \ref{fig:clusters}. Cluster 5 is in touch with clusters 1 and 2, and therefore is the last cluster to be considered.

\begin{figure}[!htb]
\begin{center}
\includegraphics[scale=0.3, angle=270]{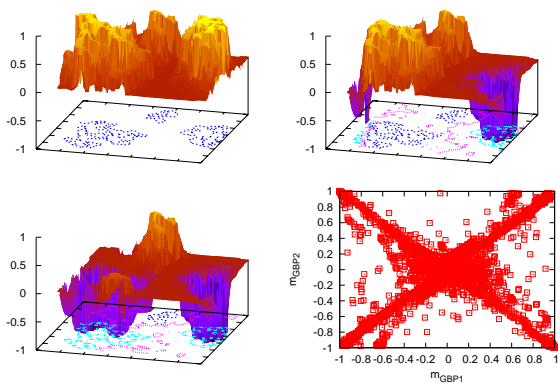} 
\end{center}
\caption{ Different GBP solutions obtained for the same $N=64\times 64$ EA2D system. The first 3 figures show the magnetization of each spin at positions $(x,y)$ in the lattice. To help the eye in recognizing the three different clusters of spins, the first GBP state is used as reference. The $z$-axis is the projection of each site magnetization onto the direction of the first solution found $m^{\beta,1}(x,y) = m^\beta(x,y) \text{Sign}(m^1(x,y))$. Three different mostly independent clusters can be seen in the contour surfaces of the top-left plot, and 3D plots show how they can switch directions from one GBP solution to other. In the bottom-right we plot $m^1_i$ {\it vs} $m^2_i$ for each spin in the system in two different GBP solutions.}
\label{fig:gbpsols}
\end{figure}

Once the clusters are identified, we use the message passing program starting from the given solution, and seek convergence after reverting the sign of all messages pointing to the spins in a given cluster. This is tantamount to reverting all magnetizations in the given cluster, while keeping the others in their original state. If we have found $n$ clusters, and all of them can flip independently, we can try convergence to $2^{n}$ solutions. Our final set of GBP states will be created out of all different GBP solutions found by this procedure. In figure \ref{fig:gbpsols} we show three different solutions obtained this way.

\subsection{One case example}
Take, for instance, the sample of figure \ref{fig:clusters}. From the possible $2^5=32$ different GBP initial conditions (including the trivial symmetry of the system), we found only $24$ solutions. To illustrate the connection between these states and Monte Carlo dynamics, we run a MC simulation with Metropolis updating rule, of the $N=128\times 128$ system and averaged the local magnetization in a time window of 1000 MC steps: 
\begin{equation}
m_i^{\small{MC}}(T=t/1000) = 1/1000 \sum_{t=T+1}^{T+1000} s_i
\end{equation}
Then we project this quantity over the GBP local magnetization of each state $\alpha$, defining the quantity
\begin{equation}
 q_{\alpha,\small{MC}}(t) = \frac 1 N \sum_i m_i^\alpha m_i^{\small{MC}}. 
\end{equation}

\begin{figure*}[!htb]
\begin{center}
\includegraphics[scale=0.28, angle=270]{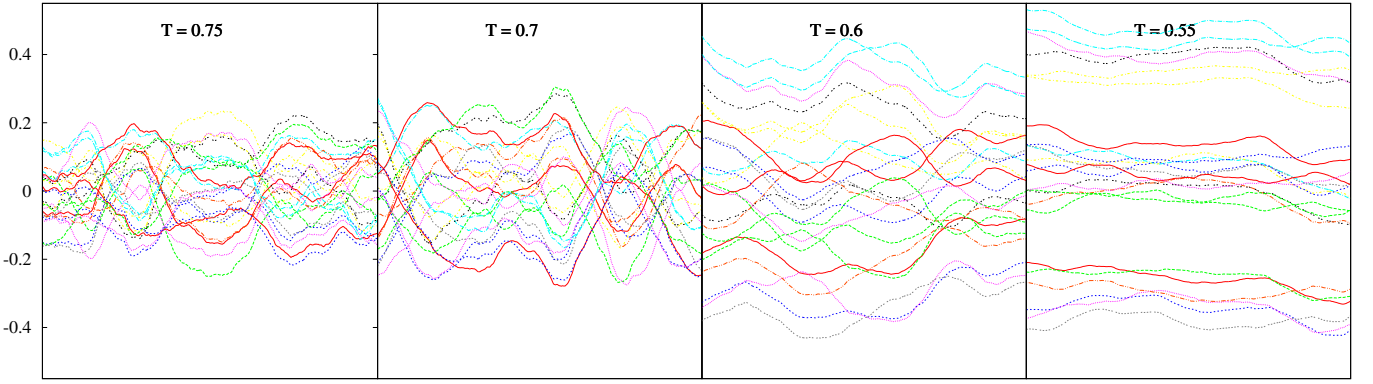} 
\end{center}
\caption{
Overlap $q_{\alpha,MC}$ between Monte Carlo magnetizations over a short
time window (1000 MCS) and the GBP predicted magnetizations, as a
function of Monte Carlo time. Each line corresponds to the projection on
a different GBP solution.  The data is smoothed with a nearest neighbor smoother, to average out high frequency Monte Carlo noise.}
\label{fig:mcprojection}
\end{figure*}

In figure \ref{fig:mcprojection} we show this projection as a function of MC-time $T$. It can be seen how the projection over the GBP states is non trivial, growing in absolute value and in time persistence as temperature goes down, and how the system switches from one state to the other, remaining most of the time nearby one of these states. 

\begin{figure}[!htb]
\begin{center}
\includegraphics[width=0.3\textwidth, angle=270]{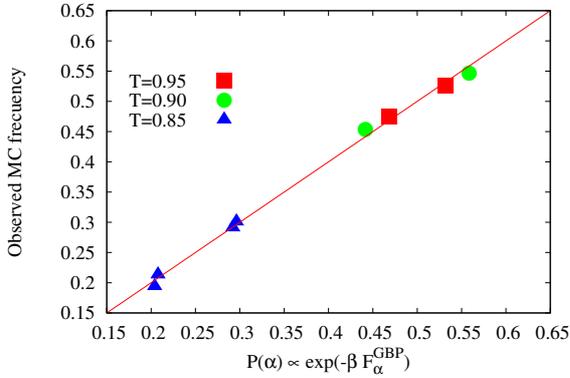} 
 \end{center}
\caption{At temperatures $T=0.95,0.9,0.85$ GBP finds two, two and three independent clusters, and therefore there are $2\times 2^1$, $2\times 2^1$ and $2\times 2^2$ GBP solutions respectively. The first factor $2$ corresponds to the natural symmetry $s_i\to-s_i$. Since symmetric solutions are equivalent in all senses, they will be taken as one solution. The observed time fraction that the Monte Carlo dynamics stays in the vicinity of each GBP solution is plotted against its predicted value from the GBP free energy in this particular 2DEA instance.}
\label{fig:freeencorr}
 \end{figure}

Furthermore, if GBP states are the metastable states, then the time that the system is nearby any GBP solution $\alpha$ should be proportional to the exponential of its free energy $F_\alpha$ that we can estimate by GBP
\begin{equation}
 T_\alpha \propto w_\alpha \:=\: \frac {\exp (-\beta F_\alpha) } {\sum_{\alpha'=1}^n \exp (-\beta F_{\alpha'}) }.
\end{equation}
In figure \ref{fig:freeencorr} this is shown to be the case for a system of $N=64\times64$ spins, at three different temperatures. We define the system to be near configuration $\alpha$ at time $t$ if its overlap is the highest :
\begin{equation}
 \forall_\gamma \:\:  q_{\gamma,\small{MC}}(t) \leq q_{\alpha,\small{MC}}(t)   .
\end{equation}
 An {\it experimental } frequency of each state is computed as the amount of Monte Carlo time $T_\alpha$ the system stays in the vicinity of GBP state $\alpha$, divided by the total Monte Carlo time of the experiment. This frequency is very well predicted by $w_\alpha$. 

\section{Monte Carlo replicas overlap}
So far we have shown that GBP locates metastable states in the Monte Carlo dynamics of the EA in 2D. Therefore the Boltzmann measure can be approximated as the linear combination of each GBP state measure
\begin{equation}
  P(\mathbf{s}) = \sum_{\alpha} w_\alpha P_\alpha(\mathbf{s}) \mathbb{I}(\mathbf{s}\in \alpha).
\end{equation}
Next we use this fact to predict some non trivial MC quantities using only GBP fixed points. 

One key parameter in disordered systems is the overlap between two different replicas of the system:
\begin{equation}
q = \frac 1 N \sum_i^N s_i^1 s_i^2. 
\end{equation}
The probability distribution of the overlap $P(q)$ provides information on the structure of states.

If each replica of the system stays close to one of the GBP states (for a time that can be estimated from the GBP free energy), we should be able to reproduce the statistics of the overlap $q$ from GBP data alone. We will consider that the random variable $q$ is given by a two steps stochastic process: the first one is the choice of the GBP states where the replicas are (see scheme in Fig. \ref{fig:confspace}), the second considers the stochastic fluctuation in the given states. Therefore the distribution of $q$ is a weighted sum of the probabilities of the random variables $\qab$, where $\alpha$ and $\beta$ are states indices,
\begin{equation} 
P(q) = \sum_{\alpha,\beta} w_\alpha w_\beta  P_{\alpha \beta}(q) \label{eq:overlap_dist}
\end{equation}
where
\begin{equation}
P_{\alpha \beta} (q) = \sum_{\mathbf{s}^1 \mathbf{s}^2}P_{\alpha}(\mathbf{s}^1) P_{\beta}(\mathbf{s}^2) \delta(q - \frac 1 N \sum_i s^1_i s^2_i)
\end{equation}
is the distribution of the overlap between two replicas when they are in states $\alpha$ and $\beta$, and  $w_\alpha w_\beta$ is the probability of such a situation.

The expected value of $\qab$ is readily given in terms of averages in the GBP states
\begin{equation} 
 \qab = \langle \frac 1 N \sum_i \sia \sib \rangle
= \frac 1 N \sum_i \mia \mib. \label{eq:mean_qab}
\end{equation}
On the other hand, the computation of the variance is harder and requires the estimation of the correlations as we show next.

\subsection{Estimating the variance}

The variance of the overlap between states $\alpha$ and $\beta$ is given by:
\begin{equation}
\sigma^2_{\alpha \beta} = \<\left(\frac 1 N \sum_i \sia \sib - \qab \right)^2\> = \<\left(\frac 1 N \sum_i \sia \sib \right)^2\> - \qab^2
\end{equation}
The first term in the rhs can be written in terms of connected correlations in a state, $C_{ij}^\alpha =\< \sia \sja \> - \mia \mja$, by
\begin{multline}
 \<\left(\frac 1 N \sum_i \sia \sib \right)^2\> = \frac 1 {N^2}  \sum_{ij} (\Cija+\mia\mja)(\Cijb+\mib\mjb) =\\
= \frac 1 {N^2}  \sum_{ij} \left( \Cija \Cijb + \mia\mja \Cijb + \mib \mjb \Cija \right) +  \qab^2
\end{multline}
From this we finally get
\begin{equation}
\sigma^2_{\alpha \beta} =  \frac 1 {N^2}  \sum_{ij} \left( \Cija \Cijb + \mia\mja \Cijb + \mib \mjb \Cija \right) \label{eq:sigmaab}
\end{equation}

Connected correlations between spins $\Cija$ can be approximated in
two ways: with a generalized susceptibility propagation algorithm, or
using fluctuation-dissipation relations withing GBP approximation. The
generalized susceptibility algorithm, though somehow intuitive, to the
best of our knowledge has not been developed so far. Instead, we can
obtain the connected correlations $\Cija$ in an {\it experimental} way
from GBP, by introducing a small external field over the spins (one at
a time) and using the fluctuation dissipation relation $\Cija =
\frac{\partial \mia }{\partial h_j}$

This procedure is a little bit more cumbersome. We need to run GBP,
and within every solution found, compute $\Cija$ for every pair of
spins in the system. This calculation requires the introduction of a
small field $\delta h_i$ over spin $i$, then running GBP some more
steps until convergence, and then computing $\Cija \simeq \delta
m_j^\alpha / \delta h_i$. Note that every time we put the probe field
$\delta h_i$ we get, after convergence, an estimate for $N$
correlations. Fortunately, for estimating $\sigma^2_{\alpha \beta}$
the correlations are averaged over all site pairs, and thus it is
enough to sample a random, and large enough, subset of the
correlations. Therefore, we have selected $50$ random spins
in the system, and run GBP with the external field
on each of them to get $50 \times N$ estimates of $\Cija$.

\begin{figure}
\begin{center}
\includegraphics[scale=0.3, angle=270]{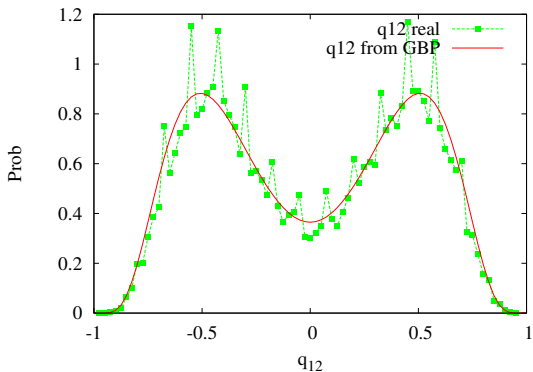} 
 \end{center}
\caption{{\bf Points} Distribution of the overlap $q_{12}$ between two independent Monte Carlo simulations of a 2DEA system (N=$64\times 64$) at temperature $T=0.65$. {\bf Solid Line} Distribution (\ref{eq:overlap_dist}) obtained from GBP states at the same temperature.}
\label{fig:overlapdist}
 \end{figure}

Given $\qab$ and $\sigma_{\alpha \beta}^2$, we would like to
approximate $P_{\alpha \beta}(q)$ by a suitable function with average
$\qab$ and variance $\sigma_{\alpha \beta}^2$.  Unfortunately a simple
Gaussian ansatz is deemed to fail because $q$ is bounded in $[-1,1]$.
We alleviate this problem by assuming normal fluctuations for the
unbounded variable $h \equiv \arctanh (q)$, which has been proved
effective in previous works \cite{manypaper}.

In figure \ref{fig:overlapdist} we show the results of our
analysis. The figure compares Monte Carlo measurements for $P(q)$ in a
system of $N=64\times 64$ spins at $T=0.65$ with function
(\ref{eq:overlap_dist}) showing a remarkable coincidence between the two.

\section{Conclusions}

We have used the plaquette-CVM approximation to the free energy, and
the corresponding Generalized Belief Propagation algorithm to study
the intermediate temperature regime of the 2D Edwards-Anderson model.
We have shown that
the spin glass solutions obtained in the temperature range $\beta \in
[0.79,1.2]$ give very useful information about the dynamics of the
actual finite-size system. Indeed, the Monte Carlo dynamic stays near
the GBP solutions a fraction of time proportional to the statistical
weigth predicted by the plaquette-CVM approximation.
Moreover the overlap distribution $P(q)$ can be well approximated from
the GBP fixed point solutions.

In our opinion this is a very promising result which may pave the way
towards a better use of CVM approximations to improve the numerical
study of complex and disordered systems.
For example, one can think of speeding up a Monte Carlo simulation by
proposing a cluster flipping move, using the clusters found by the GBP
algorithm.

Furthermore it could be interesting to study whether some specific
features of the aging dynamics, as e.g.\ the strong timescale
sepration in the flipping times at low temperatures
\cite{fede1, gleiser}, can be extracted from the GBP fixed
point solutions.

\section{ Acknoledgments}

This research has received financial support from the
Italian Research Minister through the FIRB project No. RBFR086NN1.

\bibliography{bibliografia}

\begin{thebibliography}{10}

\bibitem{aurelien}
Florent~Krzakala Aur\'elien~Decelle.
\newblock Belief-propagation guided monte-carlo sampling.
\newblock {\em http://arxiv.org/abs/1307.7846}, 2013.

\bibitem{bethe}
H.~A. Bethe.
\newblock Statistical theory of superlattices.
\newblock {\em Proc. R. Soc. A.}, 150:552--575, 1935.

\bibitem{col2}
A.~Braunstein, R.~Mulet, A.~Pagnani, M.~Weigt, and R.~Zecchina.
\newblock Polynomial iterative algorithms for coloring and analyzing random
  graphs.
\newblock {\em Phys. Rev. E}, 68(3):036702, Sep 2003.

\bibitem{GBPGF}
Eduardo Dominguez, Alejandro Lage-Castellanos, Roberto Mulet, Federico
  Ricci-Tersenghi, and Tommaso Rizzo.
\newblock Characterizing and improving {G}eneralized {B}elief {P}ropagation
  algorithms on the 2{D} {E}dwards–{A}nderson model.
\newblock {\em J. Stat. Mech.: Theor. Exp.}, 2011(12):P12007, 2011.

\bibitem{EA_original}
S~F {E}dwards and P~W {A}nderson.
\newblock Theory of spin glasses.
\newblock {\em J. Phys. F: Met. Phys.}, 5(5):965, 1975.

\bibitem{zhou3}
Chuang~Wang Haijun~Zhou.
\newblock Region graph partition function expansion and approximate free energy
  landscapes: Theory and some numerical results.
\newblock {\em Journal of Statistical Physics}, 148:513--547, 2013.

\bibitem{zhou2}
Jing-Qing~Xiao Haijun~Zhou, Chuang~Wang and Zedong Bi.
\newblock Partition function expansion on region-graphs and message-passing
  equations.
\newblock {\em J. Stat. Mech.: Theor. Exper.}, 44:L12001, 2011.

\bibitem{HAK03}
T.~Heskes, C.~A. Albers, and H.~J. Kappen.
\newblock Approximate inference and constrained optimization.
\newblock {\em UAI-03}, pages 313--320, 2003.

\bibitem{zhou}
Haijun~Zhou Jing-Qing~Xiao.
\newblock Partition function loop series for a general graphical model:
  free-energy corrections and message-passing equations.
\newblock {\em J. Phys. A: Math. Theor.}, 44:425001, 2011.

\bibitem{kikuchi}
R.~Kikuchi.
\newblock A theory of cooperative phenomena.
\newblock {\em Phys. Rev.}, 81:988, 1951.

\bibitem{dual}
Alejandro Lage-Castellanos, Roberto Mulet, Federico Ricci-Tersenghi, and
  Tommaso Rizzo.
\newblock A very fast inference algorithm for finite dimensional glasses.
\newblock {\em http://arxiv.org/abs/1102.3305}, 2011.

\bibitem{average_mulet}
Alejandro Lage-Castellanos, Roberto Mulet, Federico Ricci-Tersenghi, and
  Tommaso Rizzo.
\newblock Replica cluster variational method: {RS} solution of the 2{D}
  {E}dwards-{A}nderson model.
\newblock {\em J. Phys. A: Math. Theor.}, 46:135001, 2013.

\bibitem{MezMon}
M.~M{\'e}zard and A.~Montanari.
\newblock {\em Information, Physics, and Computation}.
\newblock Oxford Graduate Texts. OUP Oxford, 2009.

\bibitem{MP2}
Marc M{\'e}zard and G.~Parisi.
\newblock The cavity method at zero temperature.
\newblock {\em J. Stat. Phys.}, 111:1--34, Apr 2003.

\bibitem{KS2}
Marc M{\'e}zard, G.~Parisi, and R.~Zecchina.
\newblock Analytic and algorithmic solution of random satisfiability problems.
\newblock {\em Science}, 297:812--815, Aug 2002.

\bibitem{KS}
Marc M{\'e}zard and Riccardo Zecchina.
\newblock Random $k$-satisfiability problem: From an analytic solution to an
  efficient algorithm.
\newblock {\em Phys. Rev. E}, 66(5):056126, Nov 2002.

\bibitem{chertkov}
Razvan~Teodorescu Michael~Chertkov, Vladimir Y.~Chernyak.
\newblock Belief propagation and loop series on planar graphs.
\newblock {\em J. Stat. Mech.}, page P05003, 2008.

\bibitem{Col}
R.~Mulet, A.~Pagnani, M.~Weigt, and R.~Zecchina.
\newblock Coloring random graphs.
\newblock {\em Phys. Rev. Lett.}, 89(26):268701, Dec 2002.

\bibitem{Newmanbook}
M.E.J Newman and J.T. Barkema.
\newblock {\em Monte Carlo Methods in Statistical Physics}.
\newblock Oxford University Press, 2001.

\bibitem{manypaper}
R.~Alvarez~Ba\ nos, A.~Cruz, L.A. Fernandez, J.M. Gil-Narvion,
  A.~Gordillo-Guerrero, M.~Guidetti, D.~I\ niguez, F.~Mantovani A.~Maiorano,
  E.~Marinari, V.~Martin-Mayor, J.~Monforte-Garcia, A.~Mu\ noz Sudupe,
  D.~Navarro, G.~Parisi, S.~Perez-Gaviro, F.~Ricci-Tersenghi, J.J.
  Ruiz-Lorenzo, S.F. Schifano, B.~Seoane, A.~Tarancon, R.~Tripiccione, and
  D.~Yllanes.
\newblock Sample-to-sample fluctuations of the overlap distributions in the
  three-dimensional edwards-anderson spin glass.
\newblock {\em Physical Review B}, 84:174209, 2011.

\bibitem{pelizzola05}
Alessandro Pelizzola.
\newblock Cluster variation method in statistical physics and probabilistic
  graphical models.
\newblock {\em J. Phys. A}, 38:R309, 2005.

\bibitem{fede1}
F.~Ricci-Tersenghi and R.~Zecchina.
\newblock Glassy dynamics near zero temperature.
\newblock {\em Physical Review E}, 62:R7567--R7570, 2000.

\bibitem{tommaso}
Tommaso Rizzo, Alejandro Lage-Castellanos, Roberto Mulet, and Federico
  Ricci-Tersenghi.
\newblock Replica cluster variational method.
\newblock {\em J. Stat. Phys.}, 139:375--416, 2010.

\bibitem{gleiser}
F.~Rom\'a, S.~Bustingorry, and P.~M. Gleiser.
\newblock Signature of the ground-state topology in the low-temperature
  dynamics of spin glasses.
\newblock {\em Physical Review Letters}, 96:167205, 2006.

\bibitem{yedidia}
J.~Yedidia, W.~T. Freeman, and Y.~Weiss.
\newblock Constructing free energy approximations and generalized belief
  propagation algorithms.
\newblock {\em IEEE T. Inform. Theory}, 51:2282--2312, 2005.

\end{thebibliography}
\end{document}